\documentclass[reprint,prl,twocolumn,superscriptaddress,showpacs,showkeys,floatfix,preprintnumbers]{revtex4-1}

\pdfoutput=1

\usepackage{braket}
\usepackage{xargs}
\usepackage{mathtools}
\usepackage[english]{babel}
\usepackage[utf8]{inputenc}
\usepackage{amsmath,amssymb,braket,slashed,mathrsfs}
\usepackage{pifont}
\usepackage{MnSymbol}
\usepackage{graphicx}
\usepackage{tikz}
\usetikzlibrary{shapes,arrows,positioning}
\tikzset{decision/.style={diamond, draw, fill=blue!20, text width=4.5em, text badly centered, inner sep=0pt}}
\tikzset{block/.style={rectangle, draw, fill=blue!20, text width=10em, text centered, rounded corners, minimum width=3.5cm}}
\tikzset{block1/.style={rectangle, draw, fill=blue!20, text width=18.5em, text centered, rounded corners, minimum width=3.5cm}}
\tikzset{line/.style={draw, -latex, thick}}
\usepackage{color,hyperref}
\usepackage{multirow,array}

\usepackage{cleveref}
\newcommand{\ba}{\begin{eqnarray}}
\newcommand{\ea}{\end{eqnarray}}
\newcommand{\be}{\begin{equation}}
\newcommand{\ee}{\end{equation}}

\newcommand{\nn}{\nonumber}

\newcommand{\latt}{(L)} 

\newcommand{\innovation}{Collaborative Innovation Center of Quantum Matter, Beijing 100871, China}
\newcommand{\chep}{Center for High Energy Physics, Peking University, Beijing 100871, China}
\newcommand{\pkuphy}{School of Physics, Peking University, Beijing 100871,
China}
\newcommand{\KeyLab}{State Key Laboratory of Nuclear Physics and Technology,
Peking University, Beijing 100871, China}
\newcommand{\Uconn}{Department of Physics, University of Connecticut, Storrs, CT 06269, USA}
\newcommand{\RBRC}{RIKEN-BNL Research Center, Brookhaven National Laboratory, Building 510, Upton, NY 11973}

\begin{document}
\title{Lattice QCD calculation of the pion charge radius using a
model-independent method}

\author{Xu~Feng}\email{xu.feng@pku.edu.cn}\affiliation{\pkuphy}\affiliation{\innovation}\affiliation{\chep}\affiliation{\KeyLab}
\author{Yang~Fu}\affiliation{\pkuphy}
\author{Lu-Chang Jin}\email{ljin.luchang@gmail.com}\affiliation{\Uconn}\affiliation{\RBRC}
%

\date{\today}

\begin{abstract}

We use a method to calculate the hadron's charge radius without model-dependent
momentum extrapolations. The method does not require the additional quark propagator
inversions on the twisted boundary conditions or the computation of the momentum derivatives 
of quark propagators and thus is easy to implement.
We apply this method to the calculation of pion charge radius $\langle
    r_\pi^2\rangle$.
For comparison, we also determine $\langle r_\pi^2\rangle$ with the 
traditional approach of computing the slope of the form factors. The new method
produces results consistent with those from the traditional method and with
statistical errors 1.5\,-\,1.9 times smaller. 
For the four gauge ensembles at the physical pion masses, the statistical errors of 
$\langle r_\pi^2\rangle$ range from 2.1\% to 4.6\% by using $\lesssim50$
configurations. For the ensemble at
$m_\pi\approx 340$ MeV, the statistical uncertainty is even reduced to a
    sub-percent level.

\end{abstract}

\maketitle
   
{\em Introduction.} -- In particle physics, hadron is a bound state of quarks
and gluons, which are held together by the strong interaction force.
Different from a point-like particle, hadron has a rich internal
structure. One intrinsic property of a hadron is its charge radius, 
which corresponds to the spatial extent of the distribution of the hadron's charge. 
The accurate determination of the charge radius not only leaves us useful
information on
the size and the structure of the hadron, but also provides crucial precision tests of the Standard Model
at low energy. It is of special importance in resolving the proton radius
puzzle~\cite{Pohl:2010zza}, where two recent experiments report results which agree with 
the previous ones obtained by spectroscopy of muonic
hydrogen~\cite{Bezginov:2019mdi,Xiong:2019} and
represent a decisive step towards solving the puzzle for a decade.

In the theoretical study, the charge radius of the hadron is essentially
a non-perturbative quantity. It is highly appealing to have a reliable
calculation of this quantity with robust uncertainty estimate using lattice QCD.
The traditional approach for determining the charge radius on the lattice
involves the extrapolation of the expression $(F(q^2)-1)/q^2$ to zero momentum transfer $q^2=0$,
where $F(q^2)$ is the vector form factor.
The choices of the fit ansatz and fitting window would inevitably bring
systematic uncertainties
from modeling the momentum dependence of $F(q^2)$. To reduce such uncertainties, twisted
boundary conditions~\cite{Bedaque:2004kc,deDivitiis:2004kq} and momentum derivatives of quark
propagators~\cite{deDivitiis:2012vs,Hasan:2017wwt} are proposed
and used.

In this work, we use an approach to directly determine the charge radius
 without the momentum extrapolations. The method is easy to implement on the lattice
 calculation
with no requirement on twisted boundary conditions or sequential-source
propagators containing momentum derivatives. As an example, we apply the method 
to the calculation of pion's charge radius. This quantity has been determined
by various groups using the traditional
method~\cite{Brommel:2006ww,Aoki:2009qn,Wang:2018pii},
twisted boundary
conditions~\cite{Frezzotti:2008dr,Boyle:2008yd,Nguyen:2011ek,Brandt:2013dua,Aoki:2015pba,Koponen:2015tkr,Alexandrou:2017blh}
as well as the momentum extrapolations from the timelike region~\cite{Meyer:2011um,Feng:2014gba,Erben:2019nmx}.
In our study we find that the statistical 
errors can be reduced to 2.1\%\,-\,4.6\% at the physical point. At
$m_\pi\approx340$ MeV, we obtain a sub-percent statistical uncertainty 0.8\%,
which is about 6\,-\,13 times smaller than that from the previous calculations at
the similar pion
masses~\cite{Frezzotti:2008dr,Boyle:2008yd,Aoki:2009qn,Nguyen:2011ek,Brandt:2013dua,Aoki:2015pba}. 

Upon finishing this work, we note that a similar idea has been proposed by C. C.
Chang et. al. in Ref.~\cite{Bouchard:2016gmc} to calculate the proton charge radius. The earlier work 
along this direction can be traced back to mid 90's to calculate the slope of the 
Isgur-Wise function at zero-recoil~\cite{Aglietti:1994nx,Lellouch:1994zu}.\footnote{We thank C. C. Chang for raising this point to us.}
When utilizing this method, we find that it cannot be used directly 
in the calculation of the pion charge radius
as it suffers from significant finite-volume effects. 
We therefore develop the techniques to solve the problems, 
which are described in the following context.

{\em Charge radius in the continuum theory.} -- We start with a Euclidean hadronic function in the infinite volume
\be
H_\phi(x)=H_\phi(t,\vec{x})=\langle
0|\phi(t,\vec{x})J_\mu(0)|\pi(\vec{0})\rangle,
\ee
where $|\pi(\vec{0})\rangle$ is a pion initial state
carrying zero spatial momentum. $J_\mu$ is an electromagnetic vector current. $\phi$ is an interpolating operator, which can annihilate a pion state.
It can be chosen as e.g. a pseduoscalar operator $\bar{u}\gamma_5d$ or an axial vector
current $\bar{u}\gamma_\mu\gamma_5d$. In this study we use the hadronic function
$H(x)=\langle0|A_4(x)J_4(0)|\pi(\vec{0})\rangle$ with
$\phi=A_4=\bar{u}\gamma_4\gamma_5d$. 

At large time $t$, $H(x)$ is saturated by the single pion state
\be
\label{eq:ground_saturation}
H(x)\doteq H_{\pi}(x)\equiv
\int\frac{d^3\vec{p}}{(2\pi)^3}\frac{f_\pi}{2}(E+m_\pi)F_\pi(q^2)
e^{-Et-i\vec{p}\cdot\vec{x}},
\ee
where the symbol $\doteq$ denotes the omission of the excited states. 
The decay constant $f_\pi\approx130~\mathrm{MeV}$ is from PCAC relation 
$\langle0|A_4(0)|\pi(\vec{p})\rangle=Ef_\pi$ and
$E=\sqrt{m_\pi^2+\vec{p}^2}$ the
pion's energy. 
The pion form
factor $F_\pi(q^2)$ can be extracted from the matrix element
$\langle\pi(\vec{p})|J_4(0)|\pi(\vec{0})\rangle=(E+m_\pi)
F_\pi(q^2)$,
with $q^2=(E-m_\pi)^2-\vec{p}^2$.
In the Taylor expansion
\be
F_\pi(q^2)=\sum_{n=0}^\infty c_n\left(\frac{q^2}{m_\pi^2}\right)^n,
\ee
$c_0=1$ is required by the charge conservation and $c_1$ is related to the
mean-square charge radius via $c_1=\frac{m_\pi^2}{6}\langle r_\pi^2\rangle$.

The spatial Fourier transform of Eq.~(\ref{eq:ground_saturation}) yields
\be
\tilde{H}(t,\vec{p})\doteq \tilde{H}_{\pi}(t,\vec{p})\equiv
\frac{f_\pi}{2}(E+m_\pi)F_\pi(q^2)e^{-Et}.
\ee
The derivative of $\tilde{H}(t,\vec{p})$ at $|\vec{p}|^2=0$ leads to
\be
\label{eq:first_line}
D(t)\equiv m_\pi^2\frac{\partial \tilde{H}(t,\vec{p})}{\partial
|\vec{p}|^2}\bigg|_{|\vec{p}|^2=0}=-\frac{m_\pi^2}{3!}\int
d^3\vec{x}\,|\vec{x}|^2H(x),
\ee
while for $\tilde{H}_{\pi}(t,\vec{p})$ we have
\be
\label{eq:second_line}
\frac{m_\pi^2}{\tilde{H}_\pi(t,\vec{0})}\frac{\partial \tilde{H}_\pi(t,\vec{p})}{\partial |\vec{p}|^2}\bigg|_{|\vec{p}|^2=0}=\frac{1}{4}-\frac{m_\pi
t}{2}-c_1,
\ee
with $\tilde{H}_\pi(t,\vec{0})=f_\pi m_\pi e^{-m_\pi t}$.
Combining Eq.~(\ref{eq:first_line}) and (\ref{eq:second_line}), one can
determine $c_1$ using $H(x)$ as input through
\be
\label{eq:continuum_ratio}
R(t) = \frac{D(t)}{\tilde{H}(t,\vec{0})}\doteq\frac{1}{4}-\frac{m_\pi t}{2}-c_1.
\ee

{\em Charge radius on the lattice.} -- In a realistic lattice QCD
calculation with a lattice size of $\sim5$ fm, the finite volume truncation effects 
are very large as at the edge of box the integrand in Eq.~(\ref{eq:first_line}) scales as
$m_\pi^2|\vec{x}|^2\exp(-m_\pi \sqrt{\vec{x}^2+t^2})\sim0.53$ with
$\sqrt{\vec{x}^2+t^2}\approx |\vec{x}|\sim2.5$ fm.
Therefore Eqs.~(\ref{eq:first_line}) - (\ref{eq:continuum_ratio})
are too sloppy to be used in a precision calculation.
On the lattice with a size $L$ and a lattice spacing $a$, the hadronic function
$H^{\latt}(x)$ is approximated by
\be
H^{\latt}(x)\doteq H_\pi^{(L)}(x)
\equiv
\frac{1}{L^3}\sum_{\vec{p}\in\Gamma}\tilde{H}_\pi(t,\vec{p})\cos(\vec{p}\cdot\vec{x}),
\ee
where $\Gamma$ indicates a set of discrete momenta
$\vec{p}=\frac{2\pi}{L}\vec{n}$ ($\vec{n}\in\mathbb{Z}^3$) with component $p_i$
ranging from $-\frac{\pi}{a}\le p_i<\frac{\pi}{a}$.
Similar to Eq.~(\ref{eq:first_line}), we define
\be
D^{(L)}(t)\equiv
-\frac{m_\pi^{2}}{3!}\sum_{\vec{x}\in\mathbb{L}^3}|\vec{x}|^{2}H^{\latt}(x),
\ee
with $\vec{x}\in\mathbb{L}^3$ running through 
$x_i=-L/2,-L/2+a,\cdots,L/2-a$ for $i=1,2,3$.

Considering the lattice discretization, we propose to use the
lattice dispersion relation
$\hat{E}^2=\hat{m}^2+\sum_{i}\hat{p}_i^2$
with $a\hat{E}=2\sinh(aE/2)$,
$a\hat{m}=2\sinh(am_\pi/2)$ and $a\hat{p}_i=2\sin(ap_i/2)$. 
The notation $\sum_i$ indicates the summation over all spatial directions.
We further adopt the lattice-modified relations
\ba
\label{eq:lat_relation}
&&\langle 0|A_4(0)|\pi(\vec{p})\rangle=\hat{E}f_\pi,
\nn\\
&&\langle\pi(\vec{p})|J_4(0)|\pi(\vec{0})\rangle=(\hat{E}+\hat{m})F_\pi(\hat{q}^2),
\nn\\
&&\frac{1}{\hat{p}_0^2-\hat{E}^2}\bigg|_{p_0\to E}=\frac{1}{2\tilde{E}}\frac{1}{p_0-E}
\ea
with $a\hat{p}_0=2\sinh(ap_0/2)$ and $a\tilde{E}=\sinh(aE)$.
The square of momentum
transfer $\hat{q}^2$ is given by
$\hat{q}^2=(\hat{E}-\hat{m})^2-\sum_{i}\hat{p}_i^2$.

As a next step, we construct a ratio
\be
\label{eq:relation_FV}
R^{(L)}(t)=\frac{D^{(L)}(t)}{\tilde{H}^{(L)}(t,\vec{0})},
\ee
where $\tilde{H}^{(L)}(t,\vec{0})$ is defined as
\be
\tilde{H}^{(L)}(t,\vec{0})\equiv\sum_{\vec{x}\in\mathbb{L}^3}H^{\latt}(x)
\doteq \sum_{\vec{x}\in\mathbb{L}^3}H_\pi^{(L)}(x)= \tilde{H}_{\pi}(t,\vec{0}).
\ee
Note that $R^{(L)}(t)$
can be written as
\be
\label{eq:ratio}
R^{(L)}(t)\doteq \sum_{n=0}^{\infty}\beta_n^{(L)}(t)\,c_n
\ee
with the coefficients $\beta_{n}^{(L)}(t)$ known explicitly through
\ba
\label{eq:expression}
&&\beta_{n}^{(L)}(t)=-\frac{m_\pi^2}{3!}
\sum_{\vec{x}\in\mathbb{L}^3}|\vec{x}|^2I_n(x)
\nn\\
&&I_n(x)=\frac{1}{L^3}\sum_{\vec{p}\in\Gamma}\frac{\hat{E}}{\tilde{E}}\frac{\tilde{m}}{\hat{m}}\frac{\hat{E}+\hat{m}}{2\hat{m}}\left(\frac{\hat{q}^2}{m_\pi^2}\right)^n
e^{-(E-m_\pi)t}\cos(\vec{p}\cdot\vec{x}).
\nn\\
\ea
Here we have used the relations in Eq.~(\ref{eq:lat_relation}).
The value of $c_1$ can be approximated by $\left(
    R^{(L)}(t)-\beta_0^{(L)}(t)\right)/\beta_1^{(L)}(t)$. 
    In Eq.~(\ref{eq:expression}) the lattice cutoff effects from
    large $\hat{q}^2$ are safely controlled by the
    suppression of $e^{-(E-m_\pi)t}$ at sufficiently large $t$. Thus the
    continuum limit ($a\to 0$) can be taken safely. But one shall not consider
    to take the extreme limit of $t\to\infty$. In that limit, the hadronic function is dominated 
    by the pion state with zero momentum, 
    while the charge radius, as a slope of the form factor, 
        requires the information from different momenta.
    Taking very large $t$ certainly makes the analysis less interesting. 
Luckily, $t$ is only required to suppress the excited-state effects
and is not necessary to be very large.

    Note that when $a\to0$
    and $L\to\infty$, all the coefficients $\beta_n^{(L)}$ for $n\ge2$ vanish as in Eq.~(\ref{eq:continuum_ratio}).
    We can consider the contamination from $c_{n\ge2}$ terms as the systematic
    effects, which are well under control by using the fine lattice spacings and large
    volumes.
Therefore
Eq.~(\ref{eq:ratio}) provides a direct way to calculate the pion charge radius
using the lattice quantity $H^{\latt}(x)$ as input.

{\em Error reduction.} -- The hadronic function $H^{\latt}(x)$
is exponentially suppressed at large $|\vec{x}|$ and thus the lattice data
near the boundary of the box mainly contribute to the noise rather than the signal.
To reduce the statistical error,
we introduce an integral range $\xi L$ with $\xi\le\frac{\sqrt{3}}{2}$ (For
$\xi\le\frac{1}{2}$ the range has a spherical shape.) and define
\be
D_k^{(L,\xi)}(t)\equiv(-1)^k\frac{m_\pi^{2k}}{(2k+1)!}\sum_{|\vec{x}|\le
\xi L}|\vec{x}|^{2k}H^{\latt}(x),
\ee
which are related to $c_n$ through
\be
\frac{D_{k}^{(L,\xi)}(t)}{\tilde{H}^{(L)}(t,\vec{0})}\doteq
\sum_{n=0}^{\infty}\beta_{k,n}^{(L,\xi)}(t)\,c_n
\ee
with 
\be
\label{eq:coeff_beta}
\beta_{k,n}^{(L,\xi)}(t)=(-1)^k\frac{m_\pi^{2k}}{(2k+1)!}\sum_{|\vec{x}|\le
\xi L}|\vec{x}|^{2k}I_n(x).
\ee

To remove the systematic contamination from the $c_2$ term, we use both
$D_1^{(L,\xi)}(t)$ and $D_2^{(L,\xi)}(t)$ to construct the ratio $R^{(L,\xi)}(t)$
\be
\label{eq:M_L_R_def}
R^{(L,\xi)}(t)=\frac{f_1 D_1^{(L,\xi)}(t)+f_2 D_2^{(L,\xi)}(t)}{\tilde{H}^{(L)}(t,\vec{0})}+h,
\ee
where the parameters $f_1$, $f_2$ and $h$ are chosen to remove the $c_0$ and $c_2$ terms. Namely, we
impose three conditions
\be
\label{eq:condition}
\sum_{k=1,2}f_k\beta_{k,n}^{(L,\xi)}(t)=b_n,\quad\text{with }
b_0=-h,\,b_1=1,\,b_2=0.
\ee
Under these conditions $R^{(L,\xi)}(t)$ is given by
\be
R^{(L,\xi)}(t)\doteq c_1+\sum_{n=3}^\infty
\left(\sum_{k=1,2}f_k\beta_{k,n}^{(L,\xi)}(t)\right) c_n.
\ee
We do not use $D_k^{(L,\xi)}(t)$ for $k\ge3$ in our calculation as
the signal-to-noise ratio decreases as $k$ increases.
Although $R^{(L,\xi)}(t)$ still receives the contamination from $c_{n\ge3}$ terms,
we expect these effects are negligibly small. In the vector meson
dominance model, $c_n$ is given by $\left(\frac{m_\pi}{m_\rho}\right)^{2n}$
where $m_\rho$ is the rho meson mass. For $n\ge 3$, $c_n$ is estimated to be less than
$0.1\%$ of $c_1$.

{\em Correlator construction.} -- 
We use four gauge ensembles at the physical pion mass together with an additional one
at $m_\pi\approx340$ MeV,
generated by the RBC and UKQCD Collaborations using domain wall fermion~\cite{Blum:2014tka,lattice2018:robert}. The ensemble
parameters are shown in Table~\ref{tab:ensemble_parameter}.
We calculate the correlation function
$\langle A_4(x)J_4(0)\phi_\pi^\dagger(-t_\pi)\rangle$ using wall-source pion
interpolating operators $\phi_\pi^\dagger$, which
have a good overlap with the $\pi$ ground state. We find the ground-state
saturation for $t_\pi\gtrsim1$ fm. In practise the
values of $t_\pi$ are chosen conservatively as shown in Table~\ref{tab:ensemble_parameter}.

\begin{table}[htbp]
	\small
	\centering
	\begin{tabular}{cccccccc}
		\hline
        \hline
        Ensemble  & $m_\pi$ [MeV] & $L$ &  $T$ & $a^{-1}$ [GeV]&
        $N_{\text{conf}}$ & $N_r$  & $t_\pi/a$ \\
        \hline
        24D  & 141.2(4) & $24$ & $64$ & $1.015$ & 47 & 1024  & 10  \\
        32D  & 141.4(3) & $32$ & $64$ & $1.015$ & 47  & 2048 & 10 \\
        32D-fine & 143.2(3) & $32$ & $64$ & $1.378$ & 52 & 1024  & 14 \\
        48I & 139.1(3) & $48$ & $96$ & $1.730$ & 31 & 1024  & 16 \\
        24D-340 & 340.9(4) & $24$ & $64$ & $1.015$ & 36 & 1024  & 10 \\
        \hline
    \end{tabular}%
    \caption{Ensembles used in this work. For each ensemble we list the pion mass $m_\pi$, 
    the spatial and temporal extents, $L$ and $T$, 
    the inverse of lattice
    spacing $a^{-1}$, the number
    of configurations used, $N_{\text{conf}}$, the number of point-source
    light-quark propagator
    generated for each configuration, $N_r$, and the
    time separation, $t_\pi$, used for the $\pi$
    ground-state saturation.}
    \label{tab:ensemble_parameter}%
\end{table}%

For each ensemble, we use the gauge configurations, each separated by at least 10 trajectories. The number
of configurations used is listed in Table~\ref{tab:ensemble_parameter}.
We produce wall-source light-quark propagators on all time slices and 
point-source ones at $N_r$ random spacetime locations
$\{x_0\}$. The values of $N_r$ are shown in Table~\ref{tab:ensemble_parameter}.
For each configuration we perform $4\,N_r$ measurements of the correlator and
obtain an
average of
\ba
C(x;t_\pi)&=&\frac{1}{4N_r}\sum_{\{x_0\}}\langle A_4(x_0+x)J_4(x_0)\phi_\pi^\dagger(t_0-t_\pi)\rangle
\nn\\
&&+\langle A_4(x_0) J_4(x_0-x) \phi_\pi^\dagger(t_0-t-t_\pi)\rangle
\nn\\
&&+\langle\phi_\pi^\dagger(t_0+t_\pi)J_4(x_0)A_4(x_0-x)\rangle
\nn\\
&&+\langle\phi_\pi^\dagger(t_0+t+t_\pi)J_4(x_0+x)A_4(x_0)\rangle,
\ea
where $t_0$ and $t$ are the time component of $x_0$ and $x$, respectively.

The hadronic function $H^{\latt}(x)$ can be obtained from $C(x;t_\pi)$
through
\be
H^{\latt}(x)=N_\pi^{-1}Z_VZ_Ae^{m_\pi
t_\pi}\,C(x;t_\pi),
\ee
with the factor $N_\pi$ defined as
$N_\pi=\frac{1}{2m_\pi}\langle\pi(\vec{0})|\phi_\pi^\dagger|0\rangle$
and $Z_{V/A}$ the renormalization factor which converts the
local vector/axial-vector current to the conserved one. Note that the
overall factor $N_\pi^{-1}Z_VZ_A e^{m_\pi t_\pi}$ cancels out when building the
ratio $R^{(L,\xi)}(t)$.

\begin{figure}
    \centering
    \includegraphics[width=.48\textwidth]{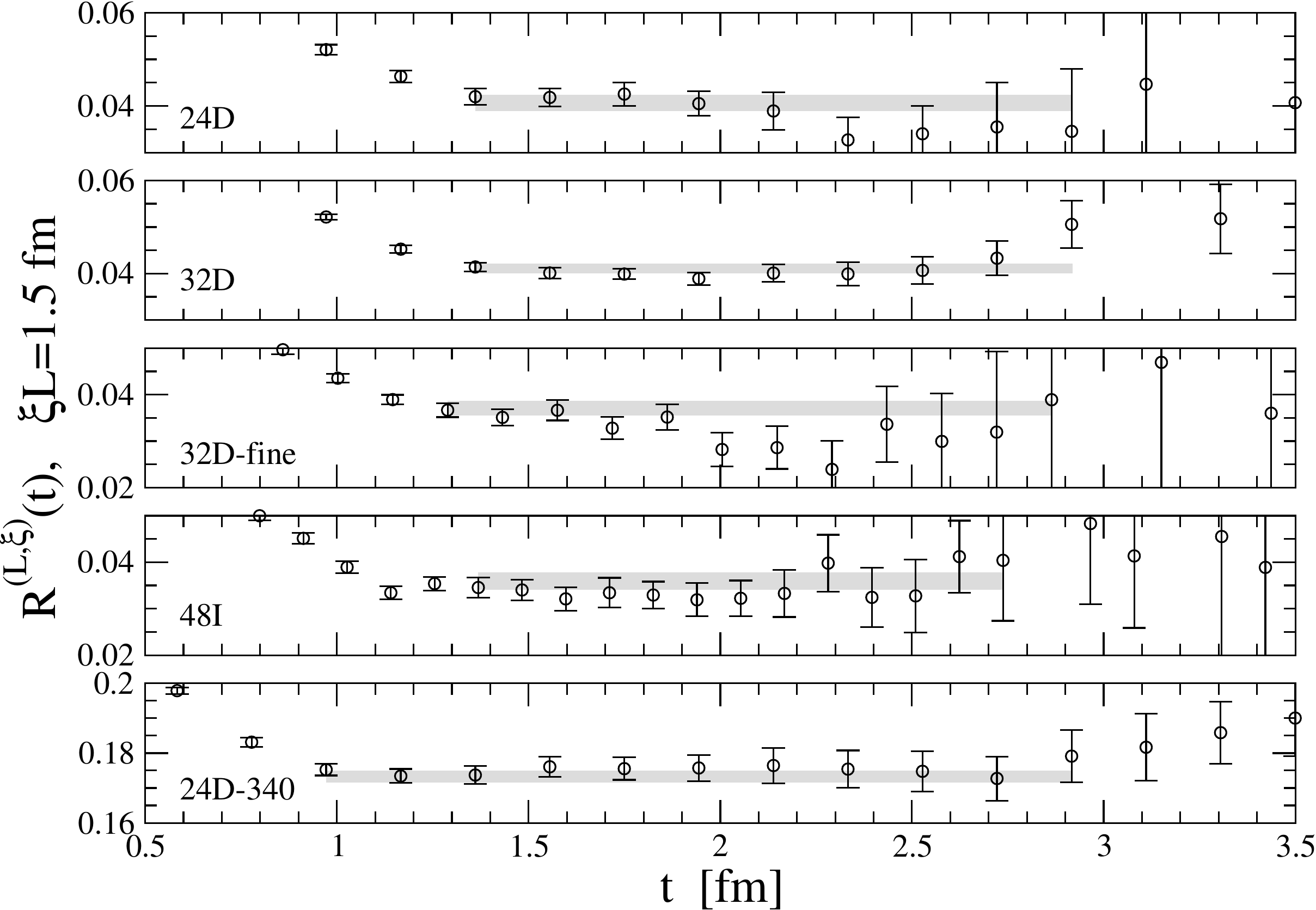}
    \caption{Results of $R^{(L,\xi)}(t)$ as a function of $t$. $R^{(L,\xi)}(t)$ are
        calculated using Eqs.~(\ref{eq:M_L_R_def}) and (\ref{eq:condition}). Here we have used
        the condition $\xi L=1.5$ fm.}
    \label{fig:R_ratio}
    \end{figure}

{\em Numerical analysis.} -- The results of $R^{(L,\xi)}(t)$ as a function of $t$ are shown in Fig.~\ref{fig:R_ratio} for
each ensemble. 
Here we have examined the $\xi L$
dependence in the lattice results and found that $\xi L=1.5$ fm is a safe choice
for the pion-ground-state dominance.
 By using $\xi L=1.5$ fm, we find that the statistical
uncertainties of $R^{(L,\xi)}(t)$ are reduced by a factor of 1.3\,-\,1.8 comparing to the
results using $\xi=\frac{\sqrt{3}}{2}$. 
We expect that the error reduction can be much more significant in the
calculation of the nucleon charge radius,
where the signal-to-noise ratio decreases as $e^{(\frac{3}{2}m_\pi-m_N)|x|}$ at
large $x$, with $m_N$ the nucleon's mass.
At large $t$, we perform a correlated fit of $R^{(L,\xi)}(t)$ to a constant and
determine $c_1$. The corresponding results for $\langle r_\pi^2\rangle$ are
listed in Table~\ref{tab:charge_radius}.

\begin{table}[htb]
\begin{tabular}{c|c|cc}
\hline\hline
\multirow{2}{*}{Ensemble}    & New & \multicolumn{2}{c}{Traditional} \\
\cline{2-4}
       &  $\langle r_\pi^2\rangle$ [fm$^2$] & $\langle
    r_\pi^2\rangle$ [fm$^2$] & $c_V$ [fm$^4$] \\
\hline
    24D & 0.476(18) & 0.466(30) & $-0.002(2)$ \\
    32D & 0.480(10) & 0.479(15) & $0.001(1)$ \\
    32D-fine & 0.423(15) & 0.409(28) & $0.001(2)$ \\
    48I & 0.434(20) & 0.395(32) & $-0.002(3)$ \\
    24D-340 & 0.3485(27) & 0.3495(44) & 0.0015(2) \\
\hline
    PDG & \multicolumn{1}{c}{0.434(5)} & \multicolumn{2}{c}{} \\
\hline
\end{tabular}
    \caption{Charge radii $\langle r_\pi^2\rangle$ from the new method by fitting
    $R^{(L,\xi)}(t)$ to a constant and from the traditional
    method by using the momentum extrapolation of $F_\pi(\hat{q}^2)$. In the
    last row, the PDG value of $\langle r_\pi^2\rangle=0.434(5)$~\cite{Tanabashi:2018oca} is listed for a
    comparison. This PDG value is obtained by combining the analysis of
    $e^+e^-\to\pi^+\pi^-$ data~\cite{Ananthanarayan:2017efc,Colangelo:2018mtw} 
    and $\pi e\to \pi e$ data~\cite{Dally:1982zk,Amendolia:1986wj,GoughEschrich:2001ji}.}
\label{tab:charge_radius}
\end{table}

    \begin{figure}
    \centering
    \includegraphics[width=.48\textwidth]{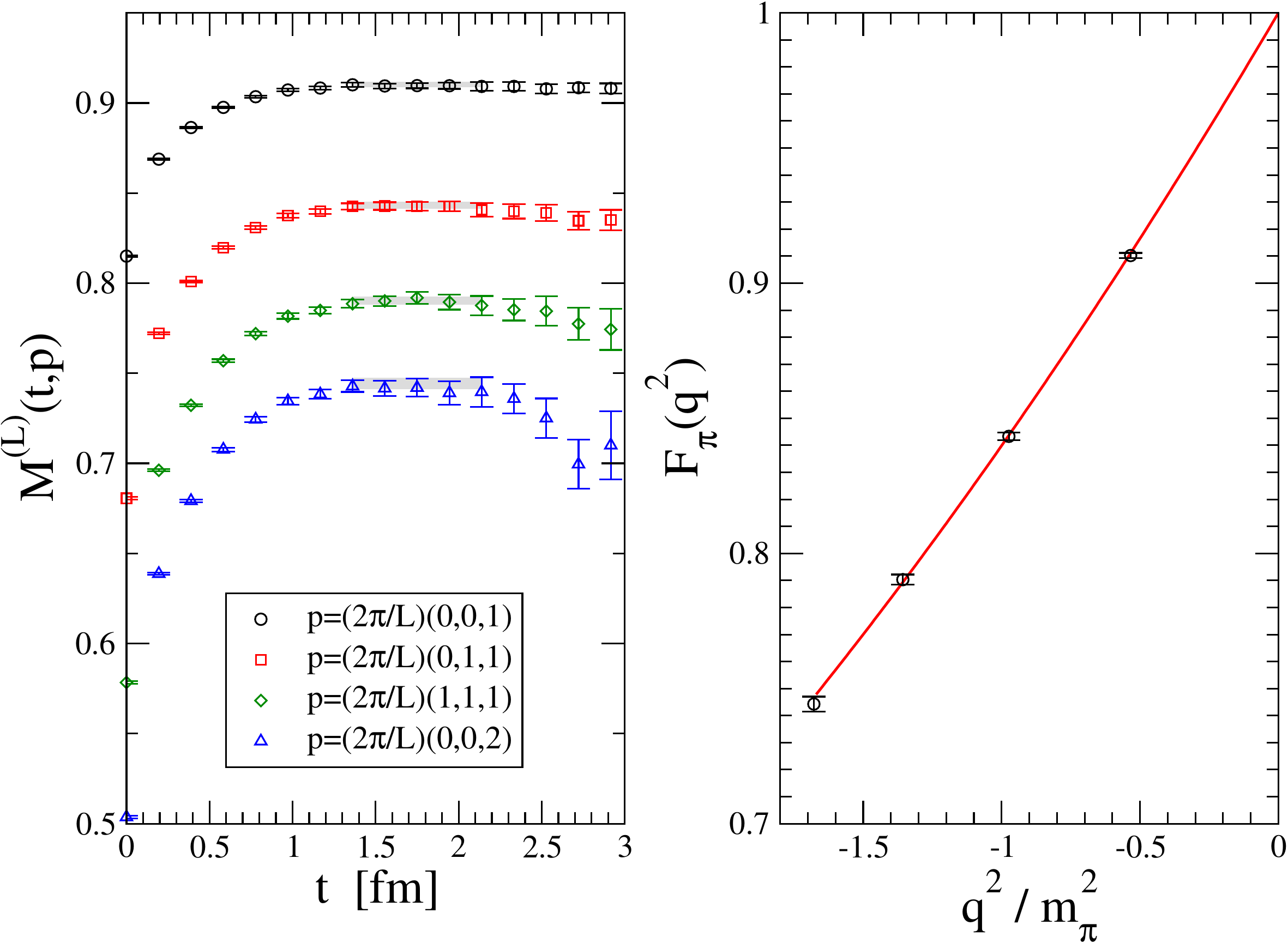}
    \caption{Using ensemble 24D-340 as an example, $M^{(L)}(t,\vec{p})$ as a function of $t$ are shown in the left
        panel and $F_\pi(\hat{q}^2)$ as a function of $\hat{q}^2/m_\pi^2$ are shown in
        the right panel.}
    \label{fig:M_ratio}
    \end{figure}

To make a comparison, we also calculate the charge radius using the tradition
method. We perform the discrete spatial Fourier transform and calculate
$\tilde{H}^{(L)}(t,\vec{p})$ using
\be
\tilde{H}^{(L)}(t,\vec{p})=\frac{1}{N_R}\sum_{\hat{R}\in
O_h}\sum_{\vec{x}\in\mathbb{L}^3}H^{\latt}(x)\cos[(\hat{R}\vec{p})\cdot\vec{x}].
\ee
where $N_R=\sum_{\hat{R}\in O_h}1$ and $O_h$ is the full cubic group for all lattice ratotations and
reflections $\hat{R}$.
We then construct the ratio
\be
M^{(L)}(t,\vec{p})=\frac{\tilde{H}^{(L)}(t,\vec{p})}{\tilde{H}^{(L)}(t,\vec{0})}
\frac{\tilde{E}}{\hat{E}}\frac{\hat{m}}{\tilde{m}}\frac{2\hat{m}}{\hat{E}+\hat{m}}e^{(E-m_\pi)t}
\doteq
F_\pi(\hat{q}^2),
\ee
with $\vec{p}=\frac{2\pi}{L}\vec{n}$ for $\vec{n}=(0,0,1)$, $(0,1,1)$, $(1,1,1)$
and $(0,0,2)$. 
Here we use the ensemble 24D-340 with smallest statitical uncertainty as an example 
and show the $t$ dependence of
$M^{(L)}(t,\vec{p})$ in the left panel of Fig.~\ref{fig:M_ratio} as well as the $\hat{q}^2$ dependence
of $F_\pi(\hat{q}^2)$ in the right panel. We perform a correlated fit of the
lattice data to a polynomial
function
\be
F_\pi(\hat{q}^2)=1+\frac{1}{6}\langle
r_\pi^2\rangle \hat{q}^2+c_V\left(\hat{q}^2\right)^2.
\ee
The fitting results are shown in Table~\ref{tab:charge_radius}. These
results are consistent with the ones from the new method, but the errors are
1.5\,-\,1.9 times larger. Therefore, we use $\langle r_\pi^2\rangle$ from the new method
in the following analysis.

{\em Systematic effects.} -- 
To examine the finite-volume effects, we use the ensembles, 24D and 32D, which
have the same
pion mass and lattice spacing but different lattice sizes, $L=4.7$ and 6.2\,fm. For these two
ensembles, the results for $\langle r_\pi^2\rangle$ are very consistent,
suggesting that the finite-volume effects are mild. This is not surprising since
the coefficients $\beta_{k,n}^{(L,\xi)}$ in Eq.~(\ref{eq:coeff_beta}) are introduced to treat
the finite-volume effects properly. 

We have four ensembles nearly at the
physical pion mass. The remaining systematic effects from the unphysical pion
mass are small and can be corrected by using the information of the fifth
ensemble, 24D-340, at $m_\pi\approx340$ MeV. We adopt the chiral extrapolation
formula~\cite{Bijnens:1998fm,Aoki:2019cca}
$\langle
r_\pi^2\rangle=\frac{1}{(4\pi
F_0)^2}\left(-\ln\frac{m_\pi^2}{m_{\pi,\text{phys}}^2}+\kappa\frac{m_\pi^2}{(4\pi
F_0)^2}+\text{const}\right)$
with the constant term including the possible lattice artifacts.
We fix $F_0=87$ MeV, a value estimated by using $F_\pi=92.2(1)$ MeV and
$F_\pi/F_0=1.062(7)$~\cite{Aoki:2019cca}, and use the ensembles 24D, 32D and 24D-340 with the same lattice spacing to study the pion mass
dependence. By extrapolating to the physical point, $\langle r_\pi^2\rangle$ for
ensembles 24D, 32D, 32D-fine and 48I are shifted by $0.2$\%, $0.3$\%, $0.5$\%,
$-0.1$\%, respectively. These changes are very small compared the statistical errors.

The largest systematic uncertainties in our study arise from the lattice
discretization effects. The values of $\langle r_\pi^2\rangle$ for 24D and 32D are 13\%
larger than that for 32D-fine, suggesting a large lattice artifact.
Unfortunately, the result from 48I cannot be used in the continuum extrapolation
together with 24D, 32D and 32D-fine ones, as the 48I ensemble 
is simulated with Iwasaki gauge action, while the other three 
use Iwasaki+DSDR action. Considering the fact that 48I has the finest lattice
spacing, we quote its value of $\langle r_\pi^2\rangle$ as the
final result and attribute to it a $\sim3$\% discretization error by an order counting 
$O((a\Lambda_{\text{QCD}})^2)$ with $\Lambda_{\text{QCD}}=300$ MeV
\be
\langle r_\pi^2\rangle=0.434(20)(13)\text{ [fm$^2$]}.
\ee
Note that the discretization error quoted here is a rough estimate. 
A further check of lattice artifacts using the finer lattice spacings is very
necessary.

{\em Conclusion.} -- We have used a model-independent method to calculate the hadron's charge
radius using lattice QCD. 
Given the hadronic function
$H^{(L)}(x)$ from lattice QCD, we propose to calculate a physical
quantity $O$ of interestes through the summation
\be
O=\sum_{\vec{x}}\omega(\vec{x},t)H^{(L)}(\vec{x},t),\quad\mbox{for large $t$}.
\ee
Here the weight function $\omega(\vec{x},t)$ is analytically known and contains all the
non-QCD information. In the calculation of the pion charge radius, where $O=\langle
r_\pi^2\rangle$, we have introduced three different weight functions $\omega(\vec{x},t)$,
which are encoded in the expressions of $R(t)$ in Eq.~(\ref{eq:continuum_ratio}), $R^{(L)}(t)$ in
Eq.~(\ref{eq:ratio}) and $R^{(L,\xi)}(t)$ in Eq.~(\ref{eq:M_L_R_def}). By choosing the appropriate
weight function, we are able to reduce both systematic and statistical
uncertainties. Such idea has been used in our earlier work on the calculation of QED self energies~\cite{Feng:2018qpx},
and can be extended to the lattice computation of various processes such as
$\pi^0\to\gamma\gamma$ decays.

In the calculation of the pion charge radius, our approach shows three peculiar features.
\begin{enumerate}
    \item Simplicity: The method does not require the additional quark propagator inversion on
        twisted boundary conditions or sequential-source propagators with
        momentum derivatives. It does not require the modeling of the
        momentum dependence of the form factor. The charge radius can be simply
        extracted from $R^{(L,\xi)}(t)$ at large time separation to avoid the excited-state
        contamination.
    \item Flexibility: In the whole calculation, it only requires the generation of
        the wall-source and point-source propagators. These propagators can be
        used to calculate other correlation functions in the future projects. 
        Besides, the hadronic function
        $\langle0|A_\mu(x)J_\nu(0)|\pi(\vec{0})\rangle$ constructed in this
        study can be used for other relevant physics processes, such as the
        radiative corrections to the pion's decay.
    \item Precision: The statistical uncertainties of $\langle r_\pi^2\rangle$ from the new method 
        are about 1.5\,-\,1.9 smaller times than that from the traditional method. We
        expect the method is more efficient in the nucleon sector where the
        hadronic function near the boundary of box contributes significant noise.
        Besides for the reduction of the statistical uncertainty, the model dependence from the
        choices of the fit ansatz is also avoided by using the new method.
\end{enumerate}

In this study, we find that the largest source of the uncertainty is from the
lattice discretization. This can be controlled by using gauge configurations with
finer lattice spacings and performing the continnum extrapolations. 
With the developments of supercomputers, technologies as
well as the new ideas and methods, we can foresee that in the near future
lattice QCD calculations can provide the determinations of $\langle
r_\pi^2\rangle$, which has the similar precision as the current PDG value or
even surpasses it.
These developments also shed the light on precise determinations of the proton charge
radius from first-principle theory
that can distinguish between the conflicting experimental values.

\section{Acknowledgments}

We gratefully acknowledge many helpful discussions with our colleagues from the
RBC-UKQCD Collaborations.
X.F. and Y.F. were supported in part by NSFC of China under Grant No. 11775002.
L.C.J. acknowledges  support  by  DOE  grant DE-SC0010339.
The computation is performed under the ALCC Program of
the US DOE on the Blue Gene/Q (BG/Q) Mira computer at the Argonne Leadership Class Facility,
a DOE Office of Science Facility supported under Contract DE-AC02-06CH11357.
The calculation is also carried out on Tianhe 3 prototype at Chinese National Supercomputer Center in Tianjin.

\bibliography{paper}

\begin{thebibliography}{35}%
\makeatletter
\providecommand \@ifxundefined [1]{%
 \@ifx{#1\undefined}
}%
\providecommand \@ifnum [1]{%
 \ifnum #1\expandafter \@firstoftwo
 \else \expandafter \@secondoftwo
 \fi
}%
\providecommand \@ifx [1]{%
 \ifx #1\expandafter \@firstoftwo
 \else \expandafter \@secondoftwo
 \fi
}%
\providecommand \natexlab [1]{#1}%
\providecommand \enquote  [1]{``#1''}%
\providecommand \bibnamefont  [1]{#1}%
\providecommand \bibfnamefont [1]{#1}%
\providecommand \citenamefont [1]{#1}%
\providecommand \href@noop [0]{\@secondoftwo}%
\providecommand \href [0]{\begingroup \@sanitize@url \@href}%
\providecommand \@href[1]{\@@startlink{#1}\@@href}%
\providecommand \@@href[1]{\endgroup#1\@@endlink}%
\providecommand \@sanitize@url [0]{\catcode `\\12\catcode `\$12\catcode
  `\&12\catcode `\#12\catcode `\^12\catcode `\_12\catcode `\%12\relax}%
\providecommand \@@startlink[1]{}%
\providecommand \@@endlink[0]{}%
\providecommand \url  [0]{\begingroup\@sanitize@url \@url }%
\providecommand \@url [1]{\endgroup\@href {#1}{\urlprefix }}%
\providecommand \urlprefix  [0]{URL }%
\providecommand \Eprint [0]{\href }%
\providecommand \doibase [0]{http://dx.doi.org/}%
\providecommand \selectlanguage [0]{\@gobble}%
\providecommand \bibinfo  [0]{\@secondoftwo}%
\providecommand \bibfield  [0]{\@secondoftwo}%
\providecommand \translation [1]{[#1]}%
\providecommand \BibitemOpen [0]{}%
\providecommand \bibitemStop [0]{}%
\providecommand \bibitemNoStop [0]{.\EOS\space}%
\providecommand \EOS [0]{\spacefactor3000\relax}%
\providecommand \BibitemShut  [1]{\csname bibitem#1\endcsname}%
\let\auto@bib@innerbib\@empty
\bibitem [{\citenamefont {Pohl}\ \emph {et~al.}(2010)\citenamefont {Pohl} \emph
  {et~al.}}]{Pohl:2010zza}%
  \BibitemOpen
  \bibfield  {author} {\bibinfo {author} {\bibfnamefont {R.}~\bibnamefont
  {Pohl}} \emph {et~al.},\ }\href {\doibase 10.1038/nature09250} {\bibfield
  {journal} {\bibinfo  {journal} {Nature}\ }\textbf {\bibinfo {volume} {466}},\
  \bibinfo {pages} {213} (\bibinfo {year} {2010})}\BibitemShut {NoStop}%
\bibitem [{\citenamefont {Bezginov}\ \emph {et~al.}(2019)\citenamefont
  {Bezginov}, \citenamefont {Valdez}, \citenamefont {Horbatsch}, \citenamefont
  {Marsman}, \citenamefont {Vutha},\ and\ \citenamefont
  {Hessels}}]{Bezginov:2019mdi}%
  \BibitemOpen
  \bibfield  {author} {\bibinfo {author} {\bibfnamefont {N.}~\bibnamefont
  {Bezginov}}, \bibinfo {author} {\bibfnamefont {T.}~\bibnamefont {Valdez}},
  \bibinfo {author} {\bibfnamefont {M.}~\bibnamefont {Horbatsch}}, \bibinfo
  {author} {\bibfnamefont {A.}~\bibnamefont {Marsman}}, \bibinfo {author}
  {\bibfnamefont {A.~C.}\ \bibnamefont {Vutha}}, \ and\ \bibinfo {author}
  {\bibfnamefont {E.~A.}\ \bibnamefont {Hessels}},\ }\href {\doibase
  10.1126/science.aau7807} {\bibfield  {journal} {\bibinfo  {journal}
  {Science}\ }\textbf {\bibinfo {volume} {365}},\ \bibinfo {pages} {1007}
  (\bibinfo {year} {2019})}\BibitemShut {NoStop}%
\bibitem [{\citenamefont {Xiong}\ \emph {et~al.}(2019)\citenamefont {Xiong},
  \citenamefont {Gasparian}, \citenamefont {Gao} \emph {et~al.}}]{Xiong:2019}%
  \BibitemOpen
  \bibfield  {author} {\bibinfo {author} {\bibfnamefont {W.}~\bibnamefont
  {Xiong}}, \bibinfo {author} {\bibfnamefont {A.}~\bibnamefont {Gasparian}},
  \bibinfo {author} {\bibfnamefont {H.}~\bibnamefont {Gao}},  \emph {et~al.},\
  }\href {\doibase 10.1038/s41586-019-1721-2} {\bibfield  {journal} {\bibinfo
  {journal} {Nature}\ }\textbf {\bibinfo {volume} {575}},\ \bibinfo {pages}
  {147} (\bibinfo {year} {2019})}\BibitemShut {NoStop}%
\bibitem [{\citenamefont {Bedaque}(2004)}]{Bedaque:2004kc}%
  \BibitemOpen
  \bibfield  {author} {\bibinfo {author} {\bibfnamefont {P.~F.}\ \bibnamefont
  {Bedaque}},\ }\href {\doibase 10.1016/j.physletb.2004.04.045} {\bibfield
  {journal} {\bibinfo  {journal} {Phys. Lett.}\ }\textbf {\bibinfo {volume}
  {B593}},\ \bibinfo {pages} {82} (\bibinfo {year} {2004})},\ \Eprint
  {http://arxiv.org/abs/nucl-th/0402051} {arXiv:nucl-th/0402051 [nucl-th]}
  \BibitemShut {NoStop}%
\bibitem [{\citenamefont {de~Divitiis}\ \emph {et~al.}(2004)\citenamefont
  {de~Divitiis}, \citenamefont {Petronzio},\ and\ \citenamefont
  {Tantalo}}]{deDivitiis:2004kq}%
  \BibitemOpen
  \bibfield  {author} {\bibinfo {author} {\bibfnamefont {G.~M.}\ \bibnamefont
  {de~Divitiis}}, \bibinfo {author} {\bibfnamefont {R.}~\bibnamefont
  {Petronzio}}, \ and\ \bibinfo {author} {\bibfnamefont {N.}~\bibnamefont
  {Tantalo}},\ }\href {\doibase 10.1016/j.physletb.2004.06.035} {\bibfield
  {journal} {\bibinfo  {journal} {Phys. Lett.}\ }\textbf {\bibinfo {volume}
  {B595}},\ \bibinfo {pages} {408} (\bibinfo {year} {2004})},\ \Eprint
  {http://arxiv.org/abs/hep-lat/0405002} {arXiv:hep-lat/0405002 [hep-lat]}
  \BibitemShut {NoStop}%
\bibitem [{\citenamefont {de~Divitiis}\ \emph {et~al.}(2012)\citenamefont
  {de~Divitiis}, \citenamefont {Petronzio},\ and\ \citenamefont
  {Tantalo}}]{deDivitiis:2012vs}%
  \BibitemOpen
  \bibfield  {author} {\bibinfo {author} {\bibfnamefont {G.~M.}\ \bibnamefont
  {de~Divitiis}}, \bibinfo {author} {\bibfnamefont {R.}~\bibnamefont
  {Petronzio}}, \ and\ \bibinfo {author} {\bibfnamefont {N.}~\bibnamefont
  {Tantalo}},\ }\href {\doibase 10.1016/j.physletb.2012.10.035} {\bibfield
  {journal} {\bibinfo  {journal} {Phys. Lett.}\ }\textbf {\bibinfo {volume}
  {B718}},\ \bibinfo {pages} {589} (\bibinfo {year} {2012})},\ \Eprint
  {http://arxiv.org/abs/1208.5914} {arXiv:1208.5914 [hep-lat]} \BibitemShut
  {NoStop}%
\bibitem [{\citenamefont {Hasan}\ \emph {et~al.}(2018)\citenamefont {Hasan},
  \citenamefont {Green}, \citenamefont {Meinel}, \citenamefont {Engelhardt},
  \citenamefont {Krieg}, \citenamefont {Negele}, \citenamefont {Pochinsky},\
  and\ \citenamefont {Syritsyn}}]{Hasan:2017wwt}%
  \BibitemOpen
  \bibfield  {author} {\bibinfo {author} {\bibfnamefont {N.}~\bibnamefont
  {Hasan}}, \bibinfo {author} {\bibfnamefont {J.}~\bibnamefont {Green}},
  \bibinfo {author} {\bibfnamefont {S.}~\bibnamefont {Meinel}}, \bibinfo
  {author} {\bibfnamefont {M.}~\bibnamefont {Engelhardt}}, \bibinfo {author}
  {\bibfnamefont {S.}~\bibnamefont {Krieg}}, \bibinfo {author} {\bibfnamefont
  {J.}~\bibnamefont {Negele}}, \bibinfo {author} {\bibfnamefont
  {A.}~\bibnamefont {Pochinsky}}, \ and\ \bibinfo {author} {\bibfnamefont
  {S.}~\bibnamefont {Syritsyn}},\ }\href {\doibase 10.1103/PhysRevD.97.034504}
  {\bibfield  {journal} {\bibinfo  {journal} {Phys. Rev.}\ }\textbf {\bibinfo
  {volume} {D97}},\ \bibinfo {pages} {034504} (\bibinfo {year} {2018})},\
  \Eprint {http://arxiv.org/abs/1711.11385} {arXiv:1711.11385 [hep-lat]}
  \BibitemShut {NoStop}%
\bibitem [{\citenamefont {Brömmel}\ \emph {et~al.}(2007)\citenamefont
  {Brömmel} \emph {et~al.}}]{Brommel:2006ww}%
  \BibitemOpen
  \bibfield  {author} {\bibinfo {author} {\bibfnamefont {D.}~\bibnamefont
  {Brömmel}} \emph {et~al.} (\bibinfo {collaboration} {QCDSF/UKQCD}),\ }\href
  {\doibase 10.1140/epjc/s10052-007-0295-6} {\bibfield  {journal} {\bibinfo
  {journal} {Eur. Phys. J.}\ }\textbf {\bibinfo {volume} {C51}},\ \bibinfo
  {pages} {335} (\bibinfo {year} {2007})},\ \Eprint
  {http://arxiv.org/abs/hep-lat/0608021} {arXiv:hep-lat/0608021 [hep-lat]}
  \BibitemShut {NoStop}%
\bibitem [{\citenamefont {Aoki}\ \emph {et~al.}(2009)\citenamefont {Aoki} \emph
  {et~al.}}]{Aoki:2009qn}%
  \BibitemOpen
  \bibfield  {author} {\bibinfo {author} {\bibfnamefont {S.}~\bibnamefont
  {Aoki}} \emph {et~al.} (\bibinfo {collaboration} {JLQCD, TWQCD}),\ }\href
  {\doibase 10.1103/PhysRevD.80.034508} {\bibfield  {journal} {\bibinfo
  {journal} {Phys. Rev.}\ }\textbf {\bibinfo {volume} {D80}},\ \bibinfo {pages}
  {034508} (\bibinfo {year} {2009})},\ \Eprint {http://arxiv.org/abs/0905.2465}
  {arXiv:0905.2465 [hep-lat]} \BibitemShut {NoStop}%
\bibitem [{\citenamefont {Wang}\ \emph {et~al.}(2018)\citenamefont {Wang},
  \citenamefont {Liang}, \citenamefont {Draper}, \citenamefont {Liu},\ and\
  \citenamefont {Yang}}]{Wang:2018pii}%
  \BibitemOpen
  \bibfield  {author} {\bibinfo {author} {\bibfnamefont {G.}~\bibnamefont
  {Wang}}, \bibinfo {author} {\bibfnamefont {J.}~\bibnamefont {Liang}},
  \bibinfo {author} {\bibfnamefont {T.}~\bibnamefont {Draper}}, \bibinfo
  {author} {\bibfnamefont {K.-F.}\ \bibnamefont {Liu}}, \ and\ \bibinfo
  {author} {\bibfnamefont {Y.-B.}\ \bibnamefont {Yang}},\ }\bibfield
  {booktitle} {\emph {\bibinfo {booktitle} {{Proceedings, 36th International
  Symposium on Lattice Field Theory (Lattice 2018): East Lansing, MI, United
  States, July 22-28, 2018}}},\ }\href {\doibase 10.22323/1.334.0127}
  {\bibfield  {journal} {\bibinfo  {journal} {PoS}\ }\textbf {\bibinfo {volume}
  {LATTICE2018}},\ \bibinfo {pages} {127} (\bibinfo {year} {2018})},\ \Eprint
  {http://arxiv.org/abs/1810.12824} {arXiv:1810.12824 [hep-lat]} \BibitemShut
  {NoStop}%
\bibitem [{\citenamefont {Frezzotti}\ \emph {et~al.}(2009)\citenamefont
  {Frezzotti}, \citenamefont {Lubicz},\ and\ \citenamefont
  {Simula}}]{Frezzotti:2008dr}%
  \BibitemOpen
  \bibfield  {author} {\bibinfo {author} {\bibfnamefont {R.}~\bibnamefont
  {Frezzotti}}, \bibinfo {author} {\bibfnamefont {V.}~\bibnamefont {Lubicz}}, \
  and\ \bibinfo {author} {\bibfnamefont {S.}~\bibnamefont {Simula}} (\bibinfo
  {collaboration} {ETM}),\ }\href {\doibase 10.1103/PhysRevD.79.074506}
  {\bibfield  {journal} {\bibinfo  {journal} {Phys. Rev.}\ }\textbf {\bibinfo
  {volume} {D79}},\ \bibinfo {pages} {074506} (\bibinfo {year} {2009})},\
  \Eprint {http://arxiv.org/abs/0812.4042} {arXiv:0812.4042 [hep-lat]}
  \BibitemShut {NoStop}%
\bibitem [{\citenamefont {Boyle}\ \emph {et~al.}(2008)\citenamefont {Boyle},
  \citenamefont {Flynn}, \citenamefont {Juttner}, \citenamefont {Kelly},
  \citenamefont {de~Lima}, \citenamefont {Maynard}, \citenamefont {Sachrajda},\
  and\ \citenamefont {Zanotti}}]{Boyle:2008yd}%
  \BibitemOpen
  \bibfield  {author} {\bibinfo {author} {\bibfnamefont {P.~A.}\ \bibnamefont
  {Boyle}}, \bibinfo {author} {\bibfnamefont {J.~M.}\ \bibnamefont {Flynn}},
  \bibinfo {author} {\bibfnamefont {A.}~\bibnamefont {Juttner}}, \bibinfo
  {author} {\bibfnamefont {C.}~\bibnamefont {Kelly}}, \bibinfo {author}
  {\bibfnamefont {H.~P.}\ \bibnamefont {de~Lima}}, \bibinfo {author}
  {\bibfnamefont {C.~M.}\ \bibnamefont {Maynard}}, \bibinfo {author}
  {\bibfnamefont {C.~T.}\ \bibnamefont {Sachrajda}}, \ and\ \bibinfo {author}
  {\bibfnamefont {J.~M.}\ \bibnamefont {Zanotti}},\ }\href {\doibase
  10.1088/1126-6708/2008/07/112} {\bibfield  {journal} {\bibinfo  {journal}
  {JHEP}\ }\textbf {\bibinfo {volume} {07}},\ \bibinfo {pages} {112} (\bibinfo
  {year} {2008})},\ \Eprint {http://arxiv.org/abs/0804.3971} {arXiv:0804.3971
  [hep-lat]} \BibitemShut {NoStop}%
\bibitem [{\citenamefont {Nguyen}\ \emph {et~al.}(2011)\citenamefont {Nguyen},
  \citenamefont {Ishikawa}, \citenamefont {Ukawa},\ and\ \citenamefont
  {Ukita}}]{Nguyen:2011ek}%
  \BibitemOpen
  \bibfield  {author} {\bibinfo {author} {\bibfnamefont {O.~H.}\ \bibnamefont
  {Nguyen}}, \bibinfo {author} {\bibfnamefont {K.-I.}\ \bibnamefont
  {Ishikawa}}, \bibinfo {author} {\bibfnamefont {A.}~\bibnamefont {Ukawa}}, \
  and\ \bibinfo {author} {\bibfnamefont {N.}~\bibnamefont {Ukita}},\ }\href
  {\doibase 10.1007/JHEP04(2011)122} {\bibfield  {journal} {\bibinfo  {journal}
  {JHEP}\ }\textbf {\bibinfo {volume} {04}},\ \bibinfo {pages} {122} (\bibinfo
  {year} {2011})},\ \Eprint {http://arxiv.org/abs/1102.3652} {arXiv:1102.3652
  [hep-lat]} \BibitemShut {NoStop}%
\bibitem [{\citenamefont {Brandt}\ \emph {et~al.}(2013)\citenamefont {Brandt},
  \citenamefont {Jüttner},\ and\ \citenamefont {Wittig}}]{Brandt:2013dua}%
  \BibitemOpen
  \bibfield  {author} {\bibinfo {author} {\bibfnamefont {B.~B.}\ \bibnamefont
  {Brandt}}, \bibinfo {author} {\bibfnamefont {A.}~\bibnamefont {Jüttner}}, \
  and\ \bibinfo {author} {\bibfnamefont {H.}~\bibnamefont {Wittig}},\ }\href
  {\doibase 10.1007/JHEP11(2013)034} {\bibfield  {journal} {\bibinfo  {journal}
  {JHEP}\ }\textbf {\bibinfo {volume} {11}},\ \bibinfo {pages} {034} (\bibinfo
  {year} {2013})},\ \Eprint {http://arxiv.org/abs/1306.2916} {arXiv:1306.2916
  [hep-lat]} \BibitemShut {NoStop}%
\bibitem [{\citenamefont {Aoki}\ \emph {et~al.}(2016)\citenamefont {Aoki},
  \citenamefont {Cossu}, \citenamefont {Feng}, \citenamefont {Hashimoto},
  \citenamefont {Kaneko}, \citenamefont {Noaki},\ and\ \citenamefont
  {Onogi}}]{Aoki:2015pba}%
  \BibitemOpen
  \bibfield  {author} {\bibinfo {author} {\bibfnamefont {S.}~\bibnamefont
  {Aoki}}, \bibinfo {author} {\bibfnamefont {G.}~\bibnamefont {Cossu}},
  \bibinfo {author} {\bibfnamefont {X.}~\bibnamefont {Feng}}, \bibinfo {author}
  {\bibfnamefont {S.}~\bibnamefont {Hashimoto}}, \bibinfo {author}
  {\bibfnamefont {T.}~\bibnamefont {Kaneko}}, \bibinfo {author} {\bibfnamefont
  {J.}~\bibnamefont {Noaki}}, \ and\ \bibinfo {author} {\bibfnamefont
  {T.}~\bibnamefont {Onogi}} (\bibinfo {collaboration} {JLQCD}),\ }\href
  {\doibase 10.1103/PhysRevD.93.034504} {\bibfield  {journal} {\bibinfo
  {journal} {Phys. Rev.}\ }\textbf {\bibinfo {volume} {D93}},\ \bibinfo {pages}
  {034504} (\bibinfo {year} {2016})},\ \Eprint
  {http://arxiv.org/abs/1510.06470} {arXiv:1510.06470 [hep-lat]} \BibitemShut
  {NoStop}%
\bibitem [{\citenamefont {Koponen}\ \emph {et~al.}(2016)\citenamefont
  {Koponen}, \citenamefont {Bursa}, \citenamefont {Davies}, \citenamefont
  {Dowdall},\ and\ \citenamefont {Lepage}}]{Koponen:2015tkr}%
  \BibitemOpen
  \bibfield  {author} {\bibinfo {author} {\bibfnamefont {J.}~\bibnamefont
  {Koponen}}, \bibinfo {author} {\bibfnamefont {F.}~\bibnamefont {Bursa}},
  \bibinfo {author} {\bibfnamefont {C.~T.~H.}\ \bibnamefont {Davies}}, \bibinfo
  {author} {\bibfnamefont {R.~J.}\ \bibnamefont {Dowdall}}, \ and\ \bibinfo
  {author} {\bibfnamefont {G.~P.}\ \bibnamefont {Lepage}},\ }\href {\doibase
  10.1103/PhysRevD.93.054503} {\bibfield  {journal} {\bibinfo  {journal} {Phys.
  Rev.}\ }\textbf {\bibinfo {volume} {D93}},\ \bibinfo {pages} {054503}
  (\bibinfo {year} {2016})},\ \Eprint {http://arxiv.org/abs/1511.07382}
  {arXiv:1511.07382 [hep-lat]} \BibitemShut {NoStop}%
\bibitem [{\citenamefont {Alexandrou}\ \emph {et~al.}(2018)\citenamefont
  {Alexandrou} \emph {et~al.}}]{Alexandrou:2017blh}%
  \BibitemOpen
  \bibfield  {author} {\bibinfo {author} {\bibfnamefont {C.}~\bibnamefont
  {Alexandrou}} \emph {et~al.} (\bibinfo {collaboration} {ETM}),\ }\href
  {\doibase 10.1103/PhysRevD.97.014508} {\bibfield  {journal} {\bibinfo
  {journal} {Phys. Rev.}\ }\textbf {\bibinfo {volume} {D97}},\ \bibinfo {pages}
  {014508} (\bibinfo {year} {2018})},\ \Eprint
  {http://arxiv.org/abs/1710.10401} {arXiv:1710.10401 [hep-lat]} \BibitemShut
  {NoStop}%
\bibitem [{\citenamefont {Meyer}(2011)}]{Meyer:2011um}%
  \BibitemOpen
  \bibfield  {author} {\bibinfo {author} {\bibfnamefont {H.~B.}\ \bibnamefont
  {Meyer}},\ }\href {\doibase 10.1103/PhysRevLett.107.072002} {\bibfield
  {journal} {\bibinfo  {journal} {Phys. Rev. Lett.}\ }\textbf {\bibinfo
  {volume} {107}},\ \bibinfo {pages} {072002} (\bibinfo {year} {2011})},\
  \Eprint {http://arxiv.org/abs/1105.1892} {arXiv:1105.1892 [hep-lat]}
  \BibitemShut {NoStop}%
\bibitem [{\citenamefont {Feng}\ \emph {et~al.}(2015)\citenamefont {Feng},
  \citenamefont {Aoki}, \citenamefont {Hashimoto},\ and\ \citenamefont
  {Kaneko}}]{Feng:2014gba}%
  \BibitemOpen
  \bibfield  {author} {\bibinfo {author} {\bibfnamefont {X.}~\bibnamefont
  {Feng}}, \bibinfo {author} {\bibfnamefont {S.}~\bibnamefont {Aoki}}, \bibinfo
  {author} {\bibfnamefont {S.}~\bibnamefont {Hashimoto}}, \ and\ \bibinfo
  {author} {\bibfnamefont {T.}~\bibnamefont {Kaneko}},\ }\href {\doibase
  10.1103/PhysRevD.91.054504} {\bibfield  {journal} {\bibinfo  {journal} {Phys.
  Rev.}\ }\textbf {\bibinfo {volume} {D91}},\ \bibinfo {pages} {054504}
  (\bibinfo {year} {2015})},\ \Eprint {http://arxiv.org/abs/1412.6319}
  {arXiv:1412.6319 [hep-lat]} \BibitemShut {NoStop}%
\bibitem [{\citenamefont {Erben}\ \emph {et~al.}(2019)\citenamefont {Erben},
  \citenamefont {Green}, \citenamefont {Mohler},\ and\ \citenamefont
  {Wittig}}]{Erben:2019nmx}%
  \BibitemOpen
  \bibfield  {author} {\bibinfo {author} {\bibfnamefont {F.}~\bibnamefont
  {Erben}}, \bibinfo {author} {\bibfnamefont {J.~R.}\ \bibnamefont {Green}},
  \bibinfo {author} {\bibfnamefont {D.}~\bibnamefont {Mohler}}, \ and\ \bibinfo
  {author} {\bibfnamefont {H.}~\bibnamefont {Wittig}},\ }\href@noop {} {\
  (\bibinfo {year} {2019})},\ \Eprint {http://arxiv.org/abs/1910.01083}
  {arXiv:1910.01083 [hep-lat]} \BibitemShut {NoStop}%
\bibitem [{\citenamefont {Bouchard}\ \emph {et~al.}(2016)\citenamefont
  {Bouchard}, \citenamefont {Chang}, \citenamefont {Orginos},\ and\
  \citenamefont {Richards}}]{Bouchard:2016gmc}%
  \BibitemOpen
  \bibfield  {author} {\bibinfo {author} {\bibfnamefont {C.}~\bibnamefont
  {Bouchard}}, \bibinfo {author} {\bibfnamefont {C.~C.}\ \bibnamefont {Chang}},
  \bibinfo {author} {\bibfnamefont {K.}~\bibnamefont {Orginos}}, \ and\
  \bibinfo {author} {\bibfnamefont {D.}~\bibnamefont {Richards}},\ }\bibfield
  {booktitle} {\emph {\bibinfo {booktitle} {{Proceedings, 34th International
  Symposium on Lattice Field Theory (Lattice 2016): Southampton, UK, July
  24-30, 2016}}},\ }\href {\doibase 10.22323/1.256.0170} {\bibfield  {journal}
  {\bibinfo  {journal} {PoS}\ }\textbf {\bibinfo {volume} {LATTICE2016}},\
  \bibinfo {pages} {170} (\bibinfo {year} {2016})},\ \Eprint
  {http://arxiv.org/abs/1610.02354} {arXiv:1610.02354 [hep-lat]} \BibitemShut
  {NoStop}%
\bibitem [{\citenamefont {Aglietti}\ \emph {et~al.}(1994)\citenamefont
  {Aglietti}, \citenamefont {Martinelli},\ and\ \citenamefont
  {Sachrajda}}]{Aglietti:1994nx}%
  \BibitemOpen
  \bibfield  {author} {\bibinfo {author} {\bibfnamefont {U.}~\bibnamefont
  {Aglietti}}, \bibinfo {author} {\bibfnamefont {G.}~\bibnamefont
  {Martinelli}}, \ and\ \bibinfo {author} {\bibfnamefont {C.~T.}\ \bibnamefont
  {Sachrajda}},\ }\href {\doibase 10.1016/0370-2693(94)00053-0} {\bibfield
  {journal} {\bibinfo  {journal} {Phys. Lett.}\ }\textbf {\bibinfo {volume}
  {B324}},\ \bibinfo {pages} {85} (\bibinfo {year} {1994})},\ \Eprint
  {http://arxiv.org/abs/hep-lat/9401004} {arXiv:hep-lat/9401004 [hep-lat]}
  \BibitemShut {NoStop}%
\bibitem [{\citenamefont {Lellouch}\ \emph {et~al.}(1995)\citenamefont
  {Lellouch}, \citenamefont {Nieves}, \citenamefont {Sachrajda}, \citenamefont
  {Stella}, \citenamefont {Wittig}, \citenamefont {Martinelli},\ and\
  \citenamefont {Richards}}]{Lellouch:1994zu}%
  \BibitemOpen
  \bibfield  {author} {\bibinfo {author} {\bibfnamefont {L.}~\bibnamefont
  {Lellouch}}, \bibinfo {author} {\bibfnamefont {J.}~\bibnamefont {Nieves}},
  \bibinfo {author} {\bibfnamefont {C.~T.}\ \bibnamefont {Sachrajda}}, \bibinfo
  {author} {\bibfnamefont {N.}~\bibnamefont {Stella}}, \bibinfo {author}
  {\bibfnamefont {H.}~\bibnamefont {Wittig}}, \bibinfo {author} {\bibfnamefont
  {G.}~\bibnamefont {Martinelli}}, \ and\ \bibinfo {author} {\bibfnamefont
  {D.~G.}\ \bibnamefont {Richards}} (\bibinfo {collaboration} {UKQCD}),\ }\href
  {\doibase 10.1016/0550-3213(95)00180-Z} {\bibfield  {journal} {\bibinfo
  {journal} {Nucl. Phys.}\ }\textbf {\bibinfo {volume} {B444}},\ \bibinfo
  {pages} {401} (\bibinfo {year} {1995})},\ \Eprint
  {http://arxiv.org/abs/hep-lat/9410013} {arXiv:hep-lat/9410013 [hep-lat]}
  \BibitemShut {NoStop}%
\bibitem [{Note1()}]{Note1}%
  \BibitemOpen
  \bibinfo {note} {We thank C. C. Chang for raising this point to
  us.}\BibitemShut {Stop}%
\bibitem [{\citenamefont {Blum}\ \emph {et~al.}(2016)\citenamefont {Blum} \emph
  {et~al.}}]{Blum:2014tka}%
  \BibitemOpen
  \bibfield  {author} {\bibinfo {author} {\bibfnamefont {T.}~\bibnamefont
  {Blum}} \emph {et~al.} (\bibinfo {collaboration} {RBC, UKQCD}),\ }\href
  {\doibase 10.1103/PhysRevD.93.074505} {\bibfield  {journal} {\bibinfo
  {journal} {Phys. Rev.}\ }\textbf {\bibinfo {volume} {D93}},\ \bibinfo {pages}
  {074505} (\bibinfo {year} {2016})},\ \Eprint {http://arxiv.org/abs/1411.7017}
  {arXiv:1411.7017 [hep-lat]} \BibitemShut {NoStop}%
\bibitem [{\citenamefont {Mawhinney}(2018)}]{lattice2018:robert}%
  \BibitemOpen
  \bibfield  {author} {\bibinfo {author} {\bibfnamefont {R.}~\bibnamefont
  {Mawhinney}},\ }in\ \href@noop {} {\emph {\bibinfo {booktitle} {{Proceedings,
  The 36th Annual International Symposium on Lattice Field Theory
  (LATTICE2018)}}}}\ (\bibinfo {year} {2018})\BibitemShut {NoStop}%
\bibitem [{\citenamefont {Tanabashi}\ \emph {et~al.}(2018)\citenamefont
  {Tanabashi} \emph {et~al.}}]{Tanabashi:2018oca}%
  \BibitemOpen
  \bibfield  {author} {\bibinfo {author} {\bibfnamefont {M.}~\bibnamefont
  {Tanabashi}} \emph {et~al.} (\bibinfo {collaboration} {Particle Data
  Group}),\ }\href {\doibase 10.1103/PhysRevD.98.030001} {\bibfield  {journal}
  {\bibinfo  {journal} {Phys. Rev.}\ }\textbf {\bibinfo {volume} {D98}},\
  \bibinfo {pages} {030001} (\bibinfo {year} {2018})}\BibitemShut {NoStop}%
\bibitem [{\citenamefont {Ananthanarayan}\ \emph {et~al.}(2017)\citenamefont
  {Ananthanarayan}, \citenamefont {Caprini},\ and\ \citenamefont
  {Das}}]{Ananthanarayan:2017efc}%
  \BibitemOpen
  \bibfield  {author} {\bibinfo {author} {\bibfnamefont {B.}~\bibnamefont
  {Ananthanarayan}}, \bibinfo {author} {\bibfnamefont {I.}~\bibnamefont
  {Caprini}}, \ and\ \bibinfo {author} {\bibfnamefont {D.}~\bibnamefont
  {Das}},\ }\href {\doibase 10.1103/PhysRevLett.119.132002} {\bibfield
  {journal} {\bibinfo  {journal} {Phys. Rev. Lett.}\ }\textbf {\bibinfo
  {volume} {119}},\ \bibinfo {pages} {132002} (\bibinfo {year} {2017})},\
  \Eprint {http://arxiv.org/abs/1706.04020} {arXiv:1706.04020 [hep-ph]}
  \BibitemShut {NoStop}%
\bibitem [{\citenamefont {Colangelo}\ \emph {et~al.}(2019)\citenamefont
  {Colangelo}, \citenamefont {Hoferichter},\ and\ \citenamefont
  {Stoffer}}]{Colangelo:2018mtw}%
  \BibitemOpen
  \bibfield  {author} {\bibinfo {author} {\bibfnamefont {G.}~\bibnamefont
  {Colangelo}}, \bibinfo {author} {\bibfnamefont {M.}~\bibnamefont
  {Hoferichter}}, \ and\ \bibinfo {author} {\bibfnamefont {P.}~\bibnamefont
  {Stoffer}},\ }\href {\doibase 10.1007/JHEP02(2019)006} {\bibfield  {journal}
  {\bibinfo  {journal} {JHEP}\ }\textbf {\bibinfo {volume} {02}},\ \bibinfo
  {pages} {006} (\bibinfo {year} {2019})},\ \Eprint
  {http://arxiv.org/abs/1810.00007} {arXiv:1810.00007 [hep-ph]} \BibitemShut
  {NoStop}%
\bibitem [{\citenamefont {Dally}\ \emph {et~al.}(1982)\citenamefont {Dally}
  \emph {et~al.}}]{Dally:1982zk}%
  \BibitemOpen
  \bibfield  {author} {\bibinfo {author} {\bibfnamefont {E.~B.}\ \bibnamefont
  {Dally}} \emph {et~al.},\ }\href {\doibase 10.1103/PhysRevLett.48.375}
  {\bibfield  {journal} {\bibinfo  {journal} {Phys. Rev. Lett.}\ }\textbf
  {\bibinfo {volume} {48}},\ \bibinfo {pages} {375} (\bibinfo {year}
  {1982})}\BibitemShut {NoStop}%
\bibitem [{\citenamefont {Amendolia}\ \emph {et~al.}(1986)\citenamefont
  {Amendolia} \emph {et~al.}}]{Amendolia:1986wj}%
  \BibitemOpen
  \bibfield  {author} {\bibinfo {author} {\bibfnamefont {S.~R.}\ \bibnamefont
  {Amendolia}} \emph {et~al.} (\bibinfo {collaboration} {NA7}),\ }\bibfield
  {booktitle} {\emph {\bibinfo {booktitle} {{Proceedings, 23RD International
  Conference on High Energy Physics, JULY 16-23, 1986, Berkeley, CA}}},\ }\href
  {\doibase 10.1016/0550-3213(86)90437-2} {\bibfield  {journal} {\bibinfo
  {journal} {Nucl. Phys.}\ }\textbf {\bibinfo {volume} {B277}},\ \bibinfo
  {pages} {168} (\bibinfo {year} {1986})}\BibitemShut {NoStop}%
\bibitem [{\citenamefont {Gough~Eschrich}\ \emph {et~al.}(2001)\citenamefont
  {Gough~Eschrich} \emph {et~al.}}]{GoughEschrich:2001ji}%
  \BibitemOpen
  \bibfield  {author} {\bibinfo {author} {\bibfnamefont {I.~M.}\ \bibnamefont
  {Gough~Eschrich}} \emph {et~al.} (\bibinfo {collaboration} {SELEX}),\ }\href
  {\doibase 10.1016/S0370-2693(01)01285-0} {\bibfield  {journal} {\bibinfo
  {journal} {Phys. Lett.}\ }\textbf {\bibinfo {volume} {B522}},\ \bibinfo
  {pages} {233} (\bibinfo {year} {2001})},\ \Eprint
  {http://arxiv.org/abs/hep-ex/0106053} {arXiv:hep-ex/0106053 [hep-ex]}
  \BibitemShut {NoStop}%
\bibitem [{\citenamefont {Bijnens}\ \emph {et~al.}(1998)\citenamefont
  {Bijnens}, \citenamefont {Colangelo},\ and\ \citenamefont
  {Talavera}}]{Bijnens:1998fm}%
  \BibitemOpen
  \bibfield  {author} {\bibinfo {author} {\bibfnamefont {J.}~\bibnamefont
  {Bijnens}}, \bibinfo {author} {\bibfnamefont {G.}~\bibnamefont {Colangelo}},
  \ and\ \bibinfo {author} {\bibfnamefont {P.}~\bibnamefont {Talavera}},\
  }\href {\doibase 10.1088/1126-6708/1998/05/014} {\bibfield  {journal}
  {\bibinfo  {journal} {JHEP}\ }\textbf {\bibinfo {volume} {05}},\ \bibinfo
  {pages} {014} (\bibinfo {year} {1998})},\ \Eprint
  {http://arxiv.org/abs/hep-ph/9805389} {arXiv:hep-ph/9805389 [hep-ph]}
  \BibitemShut {NoStop}%
\bibitem [{\citenamefont {Aoki}\ \emph {et~al.}(2019)\citenamefont {Aoki} \emph
  {et~al.}}]{Aoki:2019cca}%
  \BibitemOpen
  \bibfield  {author} {\bibinfo {author} {\bibfnamefont {S.}~\bibnamefont
  {Aoki}} \emph {et~al.} (\bibinfo {collaboration} {Flavour Lattice Averaging
  Group}),\ }\href@noop {} {\  (\bibinfo {year} {2019})},\ \Eprint
  {http://arxiv.org/abs/1902.08191} {arXiv:1902.08191 [hep-lat]} \BibitemShut
  {NoStop}%
\bibitem [{\citenamefont {Feng}\ and\ \citenamefont
  {Jin}(2019)}]{Feng:2018qpx}%
  \BibitemOpen
  \bibfield  {author} {\bibinfo {author} {\bibfnamefont {X.}~\bibnamefont
  {Feng}}\ and\ \bibinfo {author} {\bibfnamefont {L.}~\bibnamefont {Jin}},\
  }\href {\doibase 10.1103/PhysRevD.100.094509} {\bibfield  {journal} {\bibinfo
   {journal} {Phys. Rev.}\ }\textbf {\bibinfo {volume} {D100}},\ \bibinfo
  {pages} {094509} (\bibinfo {year} {2019})},\ \Eprint
  {http://arxiv.org/abs/1812.09817} {arXiv:1812.09817 [hep-lat]} \BibitemShut
  {NoStop}%
\end{thebibliography}%

\end{document}